\documentclass[12pt,onecolumn,showpacs,amssymb,aps,nofootinbib,floatfix]{revtex4-1}
 \usepackage{epsfig}
 \usepackage{xcolor}
\UseRawInputEncoding
 \newcommand{\eqref}[1]{Eq. \ref{#1} }
 \newcommand{\figref}[1]{Fig. \ref{#1} }

\newcommand{\eqcomma}{\phantom{AA},\phantom{AA}}

\begin{document}
\title{An analysis of the spin density matrix of quarkonium in heavy ion collisions}
\author{Kayman J. Gon\c{c}alves,Giorgio Torrieri}
\affiliation{Universidade Estadual de Campinas - Instituto de Fisica "Gleb Wataghin"\\
Rua Sйrgio Buarque de Holanda, 777\\
 CEP 13083-859 - Campinas SP
}
\begin{abstract}
In this addendum to \cite{GT}, we apply the techniques developed in that paper to the $J/\Psi$ and $\Upsilon$ spin alignement measurements in \cite{alice}.  We argue that while the data points to a maximally impure density matrix, consistent with Cooper-Frye/Statistical model freeze-out, a measurement of the dependence of the $\phi$-sensitive coefficients on the azimuthal angle with respect to the reaction plane would be the crucial test of this conclusion.
  \end{abstract}
\maketitle
In \cite{GT}, we have argued that vector spin alignement contains crucial information on the still-unknown spin hydrodynamic evolution in heavy ion collisions, provided that not just the $\rho_{00}$ coefficient but the ``off-diagonal'' coefficients are measured.

The former, $\rho_{00}$ was measured in \cite{alices,star} and is sensitive to $\theta$, the angle w.r.t. the spin alignement direction, which for heavy ions would be mainly the reaction plane, reflecting the vorticity structure in analogy to the global polarization measurement \cite{lisa}.

The other coefficients, related to density matrix elements $\rho_{0\pm 1,\pm 1\pm 1,\pm 1\mp 1}$ (see equation (2)  of \cite{GT}) would depend on a second ``reference'' angle $\phi$, perhaps defined in terms of the beam axis.   The lack of a straight-forward physical interpretation for these coefficients (see also \cite{xia}) perhaps lowered the priority for an invstigation of them.

But these coefficients are potentially important;  Unlike the spin $1/2$ ``qubit'' fermion, the vector meson, as a $3\times 3$ ``qutrit'' state, can be in an impure state w.r.t. any reference frame.   Therefore, deviations from a Cooper-Frye type freeze-out (similar to \cite{zanna} but with vector instead of spinor representations of SU(2)), expected when spin and vorticity are not in equilibrium \cite{uscausality}, would manifest themselves in the purity structure of the density matrix.

In \cite{GT} we illustrated this with a coalescence type model.
Coalescence of only spin within a vortical background should not change the coherence of the density matrix, since it is a unitary process and the dynamics is symmetric around the vortical axis.  But assuming vorticity and pre-existing spin density are not in equilibrium and pointing in different directions, this is no longer true \cite{GT};
Vorticity is ``classical" background, interacting with the quantum spin state, so if coalescence happens in a vortical background (i.e. if spin and vorticity are out of equilibrium) one expects impurity of the density matrix.
Mathematically, the loss of purity is manifest in Eq. (23) of \cite{GT} $P_L(w)$ representing the (unknown) classical probability of a vortex $\omega$ giving an angular momentum $L$ to the meson wavefunction.   When this probability becomes uniform ($P_L(w) \rightarrow$ constant ) we recover a maximally impure state\footnote{Through not quite the Cooper-frye ansatz of \cite{zanna}. A $P_L=$constant impure state can be regarded as a microcanonical density matrix assuming the diquark quarkonium state is exact.  The grand canonical matrix inherent in the Cooper-Frye formula of \cite{zanna} and it's vector extension would arise if all values of $L$ up to $\infty$ were allowed due to angular momentum fluctuations and a bath of degrees of freedom}.

For spin 1/2 fermions this is impossible to verify unless one knows the direction of both polarization and vorticity precisely, since the lack of purity can be ``rotated away''.   For vector bosons, impurity cannot be rotated away (see the discussion between eqs 4 and 5 of \cite{GT}) so this measurement becomes possible.
Thus, using the techniques developed in \cite{GT} the non-equilibrium between vorticity and spin density, widely expected from theory arguments (see the introduction to \cite{GT} and references therein) but quite an ``abstract'' theoretical concept, becomes experimentally observable.

While in \cite{GT} we hoped to stimulate experimental investigation in this direction, we were unaware that data already existed, not for vector mesons but for quarkonia $J/\Psi$ and $\Upsilon$ states \cite{alice}.   While of course the physics of quarkonia and vector mesons can be very different, we are looking at spin-projected density matrices.  As long as {\em hadronization} happens via the same thermal processes in light and heavy quark states, which seems at least reasonable \cite{ANDRONIC,raf,bec}, we can try to use the same techniques (although it should be noted approaches consistent with vacuum hadronization were also used \cite{pqcd1,pqcd2}).

Unlike \cite{star,alices},the ALICE Collaboration measurement of the quarkonium polarization included the off-diagonal values of the spin density matrix \cite{alice}.   Therefore, we can make a direct connection between polarization parameters $\lambda_{\theta}$,  $\lambda_{\phi}$ and $\lambda_{\theta\phi}$ and density matrix               coefficients shown in Equations 1,2 of \cite{GT}
\begin{equation}
\begin{array}{ccc}
\textcolor{black}{\rho_{00}=\frac{1+\lambda_{\theta}}{3+\lambda_{\theta}}} \eqcomma r_{1,-1}=\mathrm{Re}\left(\rho_{1,-1}\right)=\frac{\lambda_{\phi}}{3+\lambda_{\theta}} \eqcomma r_{10}=\mathrm{Re}\left(\rho_{-10}-\rho_{10} \right)=\frac{\lambda_{\theta\phi}}{3+\lambda_{\theta}}\\
\end{array}
\end{equation}
Thus, we can do the  analyses presented in \cite{GT} to relate $\lambda_{\theta,\phi,\theta\phi}$ to the wave function coherence via the parametrization in terms of Gell-mann matrices. Choosing the $n_{3-8}$ basis for this parametrization, we need to solve the following system of algebraic equations derived in\cite{GT}
\begin{eqnarray}
\frac{1}{12} \left(3 \left(n_8-\sqrt{3}\;
n_3\right) \cos \left(2 \theta _r\right)-\sqrt{3}\;
n_3+n_8+4\right)=\rho_{00} \label{sys1}\\
\frac{\left(n_8-\sqrt{3} \;n_3\right) \sin
	\left(\theta _r\right) \cos \left(\theta _r\right) \cos
	\left(\phi _r\right)}{\sqrt{2}}=r_{10} \label{sys2}\\
-\frac{\left(\sqrt{3}\;n_3+3 n_8\right) \sin
	\left(\theta _r\right) \sin \left(\phi _r\right)}{3
  \sqrt{2}}=\alpha_{10} \label{sys3}
\end{eqnarray}
\begin{equation}
\label{phirdef}
  \phi_r=-\frac{1}{2}\tan^{-1}\left(\frac{\alpha_{1,-1}}{r_{1,-1}}\right)
\end{equation}
 
Now, we will do the follow change variable $\tilde{n} = n_{8} - \sqrt{3}\; n_{3} $, and knowing that variables are equal to zero $\alpha_{10}$ and $\alpha_{1,-1}$. So, we can write this system equation the following form:

\begin{eqnarray}
\frac{1}{12} \left(3 \tilde{n}\cos \left(2 \theta _r\right)+\tilde{n}+4\right)=\rho_{00} \label{sys1}\\
\frac{\tilde{n} \sin\left(\theta _r\right) \cos \left(\theta _r\right) \cos\left(\phi _r\right)}{\sqrt{2}}=r_{10} \label{sys2}\\
\phi_r = 0 \label{phirdef}\\
\end{eqnarray}
Therefore, we have the following system solution: 
 
\begin{eqnarray}
	\tilde{n}\left(\rho_{00},r_{10}\right) =-\frac{(1-3 \rho_ {00})^2+3
   \sqrt{(1-3 \rho_{00})^4+4 (1-3
   \rho_{00})^2 r_{10}^2}}{6 \rho_{00}-2}\label{ntilde}\\
   \Theta_1\left(\rho_{00},r_{10}\right) = -\sqrt{\frac{2 (1-3 \rho_{00})^2-2 \sqrt{(1-3 \rho_{00})^4+4
   (1-3 \rho_{00})^2 r_{10}^2}+6
   r_{10}^2}{2 (1-3 \rho_{00})^2+9
   r_{10}^2}} \\ 
   \Theta_2\left(\rho_{00},r_{10}\right) = \sqrt{\frac{(1-3 \rho_{00})^2-\sqrt{(1-3 \rho_{00})^4+4 (1-3
   \rho_{00})^2 r_{10}^2}+3 r_{10}^2}{2
   (1-3 \rho_{00})^2+9 r_{10}^2}}\\
   \Theta_3\left(\rho_{00},r_{10}\right) = \frac{ \Theta_2\left(\rho_{00},r_{10}\right)\left(2
   (1-3 \rho_{00})^2+2 \sqrt{(1-3\rho_{00})^4+4 (1-3 \rho_{00})^2 r_{10}^2}\right)}{2 (3 \rho_{00}-1) r_{10}}\\
   \theta_r = \tan ^{-1}\left(\Theta_1\left(\rho_{00},r_{10}\right), \Theta_3\left(\rho_{00},r_{10}\right)\right)
\end{eqnarray}
So, using polarization parameters that were obtained from ALICE collaboration \cite{alice} to differents transversal momentum $p_{T}$ ranges, we can determine whether the density matrix represents a coherence state or not. To make it we will use the equation \ref{ntilde} and reach the following figure \ref{states}.
\begin{figure}[t]
		\begin{center}
			\epsfig{width=0.7\textwidth,figure=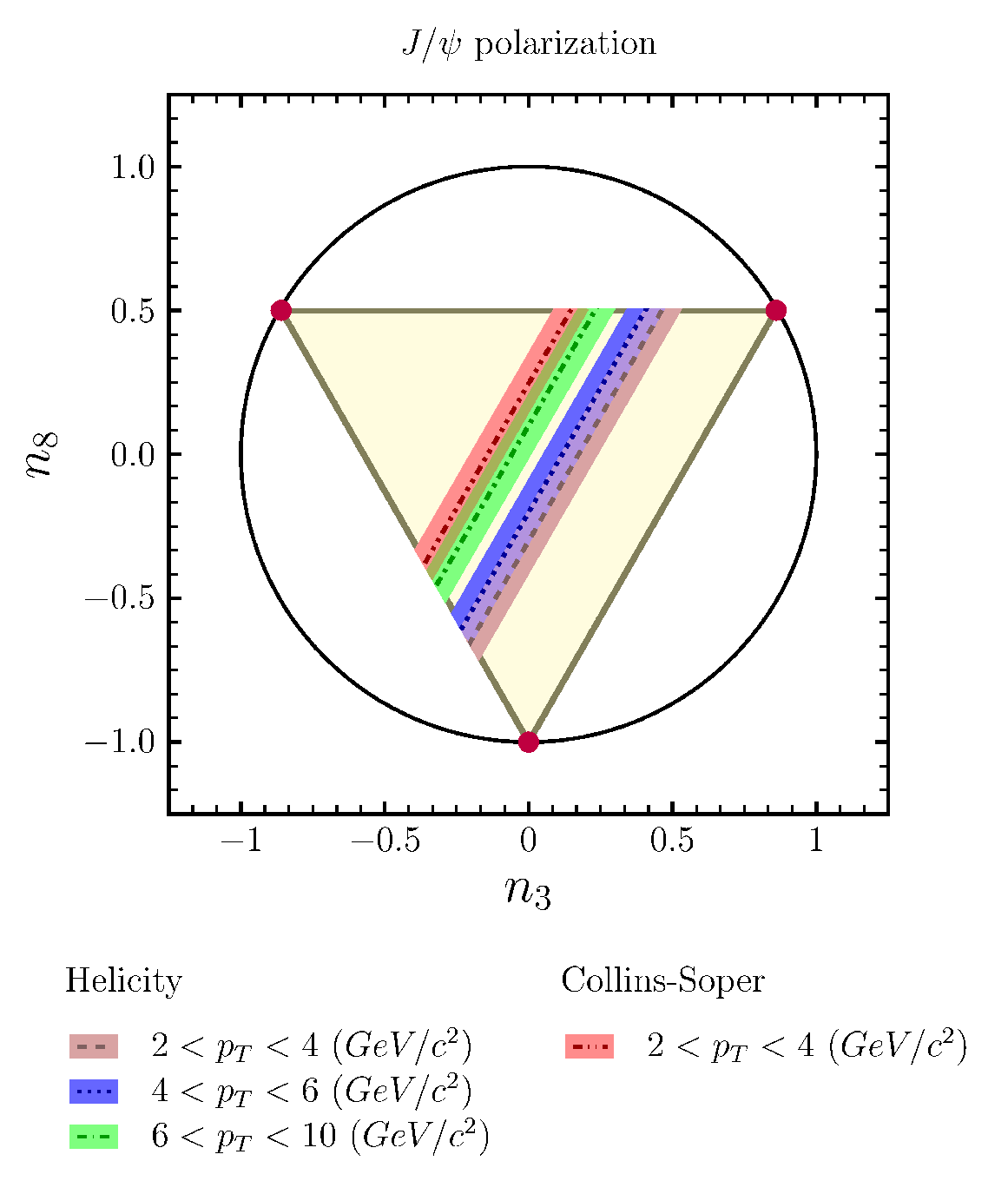}
			
\end{center}
                \caption{The experimental results of $J/\Psi$ polarization measurements analyzed in terms of the Gell-mann matrix representation of \cite{GT}.
                  These results were made using the helicity and Collins-Soper frame to $J/\psi$ polarization \cite{alice}. The other Collins-Soper values was not take in this plot because they were above the Helicity $6 < p_T < 10 ~ (GeV/C^2)$ uncertainty band as were shown in the text.   \label{states}}
	\end{figure}

Now, the coefficients in the frame Collins-Soper frame, given in the ranges $4 < p_T < 6 ~ (GeV/C^2)$ and $6 < p_T < 10 ~ (GeV/C^2)$ respectively result in $\tilde{n}=(0.09\pm 0.10, 0.02\pm 0. 07)$. Maaking the comparison with the Helicity frame value $\tilde{n}=0.09\pm 0.11$ to $6 < p_T < 10 ~ (GeV/C^2)$. Therefore,  we can see that they are the same within error bar.

Looking at the figure \ref{states}, we can conclude that the density matrix from $J/\psi$ particle does not represent the pure state since none of the values for $n_{3,8}$ obtained from the data intersects the black pink points, i.e. these points represent the pure state in other words when the density matrix satisfy $\rho^2 = \rho$. This might indicate that statistical freeze-out advocated in \cite{zanna,ANDRONIC} is a good estimate of particle production in heavy-ion collisions because the density matrix does not represent a coherent state as argued in \cite{GT}.
\begin{figure}[t]
		\begin{center}
			\epsfig{width=0.7\textwidth,figure=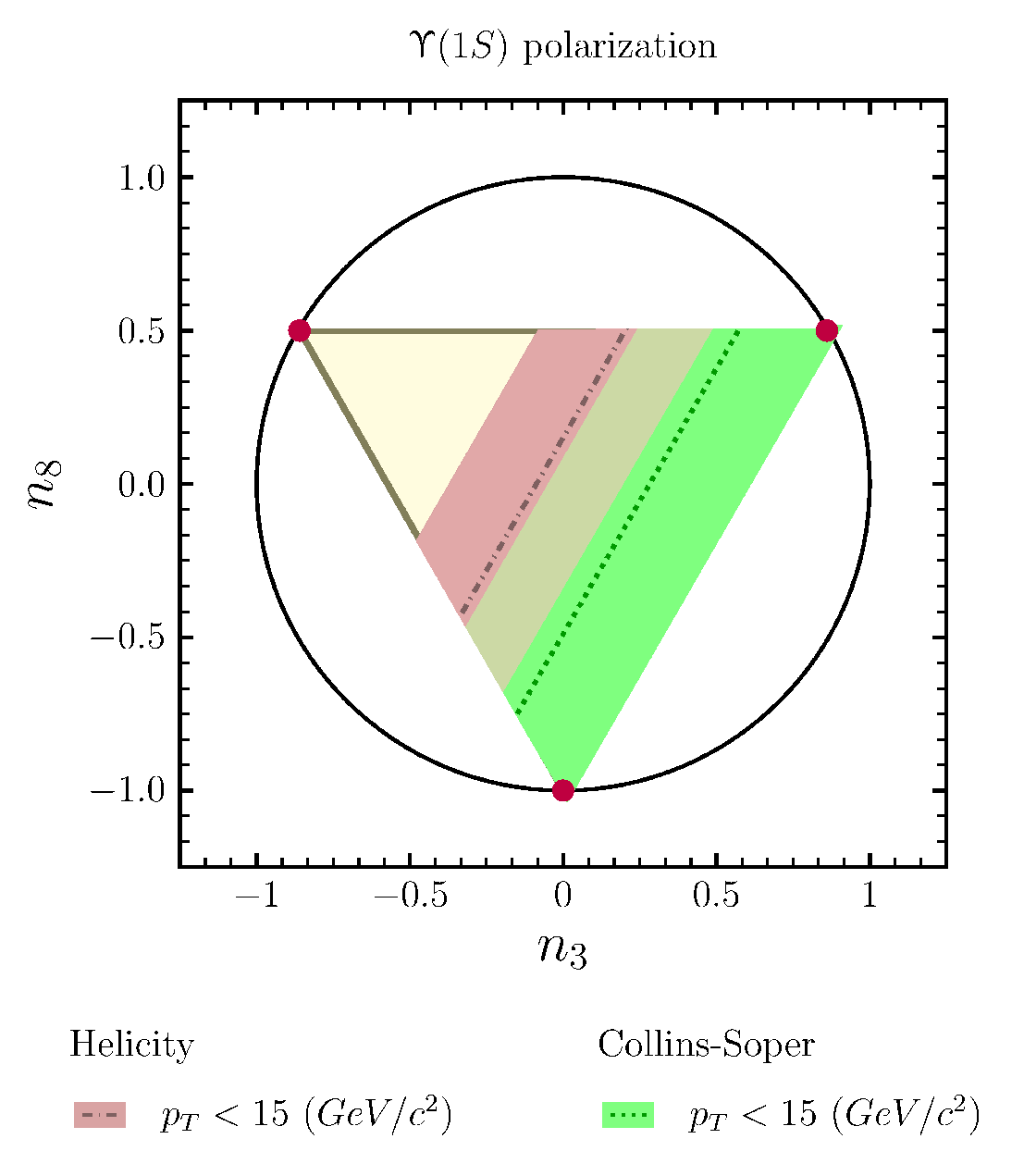}
			\label{bottomonium}
            \caption{The experimental results of $\Upsilon$ polarization measurements analyzed in terms of the Gell-mann matrix representation of \cite{GT}.These results were obtained from two different frames Helicity and Collins-Soper from $\Upsilon (1S)$ polarization.}
        \end{center}
\end{figure}
In the bottomonium $\Upsilon (1S)$ case, we can see in figure \ref{bottomonium} that because of large uncertainty do not know whether for these particles the density matrix represents a coherence state or not. 

However, taking these conclusions as definitive would be highly premature as Fig. 4 of \cite{GT} and the definition of $P_L(w)$ given in the introduction shows.
We have no idea what $P_L(w)$ is beyond the fact that it overall conserves momentum, but it acts as a projector.   One recovers a pure state when it is a $\delta_{Ll}$, (there is a certainty of it giving momentum) and a maximally mixed state when it is independent of $L$. So the measurement in Fig. 4 \cite{GT} is directly connected to how out of equilibrium vorticity and spin are, and how much vorticity vs pre-existing spin influences the final spin of the vector meson.  Linear combination of the different $L$-values in Fig. 4 of \cite{GT} are possible, illustrating a probability of different spin configurations.

\begin{figure}[t]
		\begin{center}
			\epsfig{width=0.9\textwidth,figure=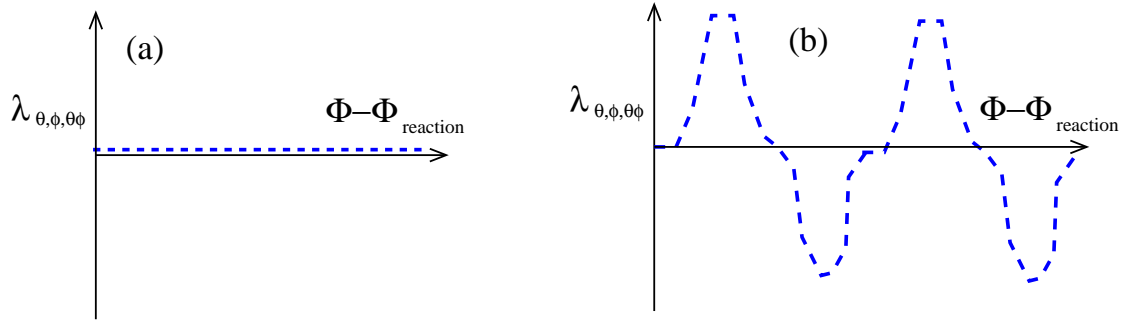}
			\label{figpl}
            \caption{\label{pl} A schematic illustration of how we expect the coefficients $\lambda_{\theta,\phi,\theta phi}$ to evolve with reaction plane angle $\Phi-\Phi_{reaction}$ in the two scenarios.   In \cite{alice}, these coefficients are compatible with zero when integrated over this angle.   A flat dependence (a) indicates a maximally incoherent Cooper-Frye/thermal type production.  A harmonic dependence (b) (whose exast shape will be sensitive to the dynamics) might indicate a more non-trivial coalescence scenario involvin vorticity.  Note that both cases are consisten with zero when integrated over $\Phi$, as was done in \cite{alice}}
        \end{center}
\end{figure}

Note that these coefficients are given in terms of an angle $\theta$, which in \cite{GT} is related to $\theta_r$, the angle between the hadronization frame and the lab frame.   This angle of course depends on the detailed hydrodynamical and spin-hydrodynamical evolution of the system, but it is obviously highly dependent on the reaction plane angle $\Phi$.   Considering the Harmonic behavior of the coefficients in Fig. 4 of \cite{GT} w.r.t. $\theta_r$ (most coefficients average to zero for all angles), therefore, it would be crucial to measure $\lambda_{\theta,\phi,\theta\phi}$ not as a function of $p_T$ as in \cite{alice} but as function of azimuthal reaction plane angle.    A modulated behavior would be a clear signature for a non-trivial $P_L(w)$ which can then be harmonically decomposed into \textcolor{black}{$L=0,1,2$} components of FIg 4 of \cite{GT} to obtain information of the impact of spin vs vorticity in $J/\Psi$ vs $\Upsilon$ hadronization.   If the dependence w.r.t. $\theta_r$ will be compatible with zero as it was for $p_T$ in each $\Phi$ bin, this is good evidence for a statistical Cooper-Frye freeze-out as in \cite{zanna,ANDRONIC}.   Schematically, these two alternatives are illustrated in \figref{pl}.

Summarizing, we have used the techniques developed in \cite{GT} on the experimental Quarkonium polarization measurement in Pb-Pb collisions.   We found the results compatible with an incoherent state,consistent with statistical production, but a measurement of modulation (or the absence of it) w.r.t. the reaction plane axis is the crucial test of this conclusion.
We are eagerly waiting for such experimental data, and also for this analysis to be extended to $K^*$ and $\phi$ vector mesons analyzed in \cite{alices,star}.

We thank Luca Micheletti for discussions at the Quark Matter conference, and Gabriel Rocha and Leticia Palhares for discussions at the Brazilian RETINHA meeting.
GT thanks CNPQ bolsa de produtividade 306152/2020-7, bolsa FAPESP 2021/01700-2 and participation 
in tematic FAPESP, 2017/05685-2. K.J.G. is supported by CAPES doctoral fellowship
88887.464061/2019-00

\end{document}